\documentclass[prl,reprint,aps,amsmath,amssymb,twocolumn,superscriptaddress,longbibliography]{revtex4-1}
\usepackage{graphicx,bm,color,appendix}
\usepackage[colorlinks,bookmarks=true,citecolor=blue,linkcolor=blue,urlcolor=blue]{hyperref}
\usepackage{times}
\usepackage{amsmath,amssymb}
\usepackage{comment}
\usepackage{appendix}

\graphicspath{{./fig/}}

\usepackage{xcolor}
\definecolor{orange}{rgb}{1,0.5,0}

\usepackage[normalem]{ulem}
\usepackage{comment}

\begin{document}
\title{Entropic $F$-function of 3D Ising conformal field theory via the fuzzy sphere regularization
}

\author{Liangdong Hu}
\affiliation{School of Sciences, Westlake University, Hangzhou 310030, China }

\author{W. Zhu}
\email{zhuwei@westlake.edu.cn}
\affiliation{School of Sciences, Westlake University, Hangzhou 310030, China }
\affiliation{Westlake Institute of Advanced Study, Westlake University, Hangzhou 310024, China }

\author{Yin-Chen He}
\email{yhe@perimeterinstitute.ca}
\affiliation{Perimeter Institute for Theoretical Physics, Waterloo, Ontario N2L 2Y5, Canada}

\begin{abstract}
The $F$-function, the three-dimensional counterpart of the central charge in the 2D conformal field theory, measures the effective number of degrees of freedom in 3D quantum field theory, and it is monotonically decreasing under the renormalization group flow. However, unlike the 2D central charge, the $F$-function is a non-local quantity and cannot be computed using correlators of local operators. Utilizing the recently proposed fuzzy sphere regularization, we have performed the first non-perturbative computation of the $F$-function for the paradigmatic 3D Ising conformal field theory through entanglement entropy. Our estimate yields $F_{\text{Ising}} \approx 0.0612(5)$, slightly smaller than the $F$-function of a real free scalar, $F_{\text{free}} = \frac{\log 2}{8} - \frac{3\zeta(3)}{16\pi^2} \approx 0.0638$, consistent with the $F$-theorem, and close to the $4-\epsilon$ expansion estimates of $F_{\text{Ising}} \approx 0.0610 \sim 0.0623$.
\end{abstract}

\maketitle

Renormalization group (RG) theory is central in theoretical physics, particularly for understanding scale-dependent behaviors in critical phenomena and quantum field theory (QFT) \cite{WilsonRG}. RG involves transformations that integrate out short-distance degrees of freedom~\cite{KadanoffBlockSpin,WilsonFisher}, revealing how physical properties evolve across different length or energy scales. A fundamental feature of RG is its inherent irreversibility, akin to the second law of thermodynamics. Specifically, under RG transformations, certain complexity measures—such as the number of degrees of freedom—will monotonically decrease~\cite{Zamolodchikov1986,CARDY1988,Osborn1989Derivation,JACKOsborn1990Analogs,Casini2004,Komargodski2011,Komargodski2012,Jafferis2012superconformal,Jafferis2011,Closset2012Contact,Myers2010Seeing,Myers2011,Casini2011,Klebanov2011,Casini2012,Liu2013,Hartman2023Averaged, Affleck1991Degeneracy,Daniel2004Entropy,Casini2016gTheorem,Cuomo2022Flow,Casini2023gTheorem,Jensen_2016_constraint,Pufu2017,Nishioka2018}. The RG irreversibility theorem provides crucial insights into deciphering the landscape of RG fixed points in QFTs, thereby elucidating the intrinsic structures of phases and phase transitions in physical systems.

The first established RG irreversibility theorem is Zamolodchikov's 
$c$-theorem in 2D QFT \cite{Zamolodchikov1986}. The $c$-theorem states that a $c$-function, associated with the central charge of 2D conformal field theories (CFTs) at RG fixed points, will monotonically decrease under RG \cite{Zamolodchikov1986}. The first generalization of the $c$-theorem in 2D is the $a$-theorem in 4D, which was conjectured by Cardy \cite{CARDY1988} and later proved by Komargodski and Schwimmer \cite{Komargodski2011}. A common feature shared by the $c$-function and $a$-function is that both are related to the conformal anomaly in curved space. However, conformal anomaly is only present in even spacetime dimensions, which poses a challenge for establishing RG irreversibility theorems in odd spacetime dimensions, particularly in 3D, which are relevant to many interesting phase transitions in condensed matter systems. Notably, the developments in supersymmetric gauge theories~\cite{Jafferis2012superconformal,Jafferis2011,Closset2012Contact} and holography~\cite{Myers2010Seeing,Myers2011} independently led to the $F$-theorem in 3D.

There are two equivalent definitions for the $F$-function at the RG fixed point. The first definition is on the three-sphere $S^3$ with a radius $r$, where the $F$-function is the universal non-divergent term of the partition function, $\log Z_{S^3} \sim \alpha_1 r^3 + \alpha_2 r - F$ \cite{Jafferis2011}. The second definition applies to $2+1$D QFT on $\mathbb R^3$ and considers the entanglement entropy (EE) of a radius $R_{\text{disk}}$ disk on flat space $\mathbb R^2$~\cite{Myers2010Seeing,Myers2011,Casini2011},
\begin{equation}\label{eq:entropicF}
S_A (R_{\text{disk}}) = -\textrm{Tr}(\rho_A \ln \rho_A) = \alpha \frac{R_{\text{disk}}}{\delta} - F.
\end{equation}
The first term is the conventional entanglement area law, with $\delta$ being the UV regulator, such as the lattice spacing. The sub-leading term is the $F$-function, which is universal at the RG fixed point (e.g., $R_{\text{disk}}\gg \delta$). One can remove the UV divergent entanglement area law term by defining a renormalized entanglement entropy (REE)~\cite{Liu2013,Casini2012},
\begin{equation}\label{eq:REE}
\mathcal F(R_{\text{disk}}) = (R_{\text{disk}} \partial_{R_{\text{disk}}} -1) S_A(R_{\text{disk}}),
\end{equation}
which becomes the $F$-function at the RG fixed point when $R_{\text{disk}}/\delta\rightarrow \infty$. It was proven by Casini and Huerta that REE is RG monotonic \cite{Casini2012}, $\mathcal F' \le 0$, hence establishing the $F$-theorem, i.e., $F_{UV}>F_{IR}$. The $F$-theorem and the value of the $F$-function at fixed points provide valuable insights for understanding QFTs, as they constrain the RG flow between different fixed points. For example, the $F$-theorem has been proposed as a tool for determining the conformal window of critical gauge theories~\cite{Grover2014Entanglement,Giombi2015QEDF}, an open problem of general interest to both the high-energy and condensed matter communities.

\begin{figure}[b]
    \centering
    \includegraphics[width=0.49\textwidth]{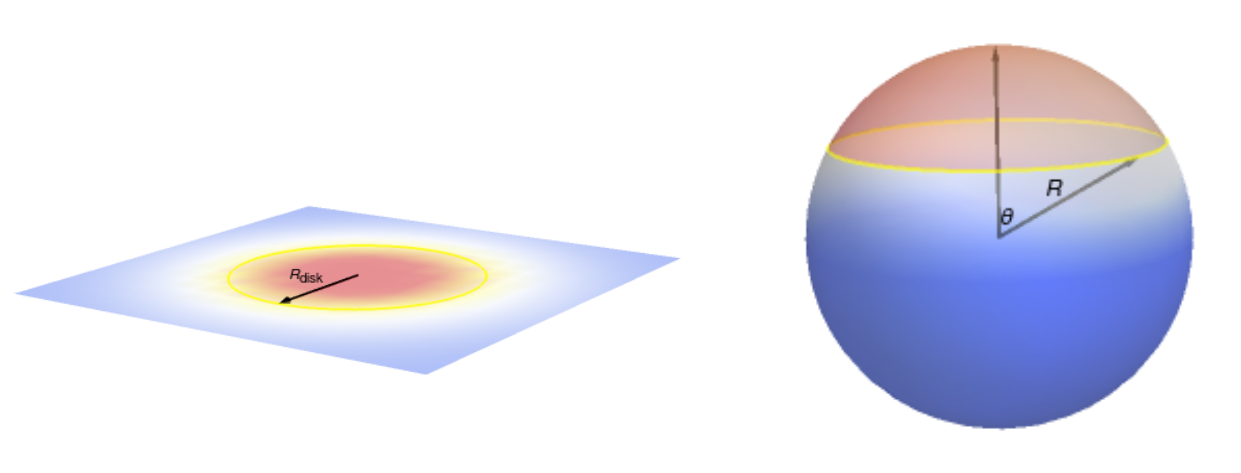}
    \caption{Entanglement bipartition on the 2D flat plane (left) and 2D curved sphere (right), where the yellow line represents the entanglement bipartition cut in the real-space.} 
    \label{fig:illustration}
\end{figure}

Although the $F$-function is a fundamental quantity characterizing 3D CFTs, so far there is no non-perturbative computation of it for any interacting CFT without supersymmetry, including the prominent example of the 3D Ising CFT. A major challenge with the $F$-function is its non-local nature, which precludes its extraction through correlation functions of local operators, unlike the 2D $c$-function and 4D $a$-function. One manifestation of the non-local nature of $F$ is that it can be nontrivial at a non-conformal fixed point. Indeed, for a topologically ordered phase described by topological quantum field theory, $F$ equals to the topological entanglement entropy (TEE) \cite{Levin2006,Kitaev2006}. However, conventional computation schemes for TEE do not apply to $F$ of a CFT due to several complications arising from the gaplessness of the CFT. Firstly, the entanglement cut must be smooth without corners; otherwise, corner contributions~\cite{Bueno2015} will contaminate $F$. Secondly, when computing EE on a torus geometry ($T^2\times \mathbb R$), as most lattice model simulations do, the perimeter $l$ of the entanglement cut should be significantly smaller than the system size $L$. This is because a torus is not conformally equivalent to $R^3$, where $F$ is originally defined. It means that cutting the torus into two halves will not yield $F$~\footnote{This results in another universal quantity different from $F$, which has been shown to be RG non-monotonic~\cite{Whitsitt2017Entanglement}.}, even though the entanglement cut is smooth. Thirdly, the definition of $F$ in Eq.~\eqref{eq:entropicF} is valid only for EE, not for the Rényi entropy~\cite{Klebanov2012}--the entanglement quantity accessible to quantum Monte Carlo simulations~\cite{Melko2010Renyi}.

Recently, a novel scheme known as fuzzy sphere regularization has been proposed for studying 3D CFTs \cite{ZHHHH2022}. The basic idea involves utilizing spherical Landau levels to realize 3D CFTs on the spacetime cylinder $S^2\times \mathbb R$. It demonstrates surprising efficiency in computing conformal data~\cite{hu2023operator,Han2023Conformal} and broad scope for generalization to other CFTs, such as $O(N)$ Wilson-Fisher~\cite{han2023o3}, critical gauge theories \cite{zhou2023so5} and defect CFTs \cite{hu2023defect,Zhou2024gfunction}. For the $F$-function concerned here, the fuzzy sphere offers a number of unique advantages in computing it. Firstly, the fuzzy sphere model exhibits a very small finite-size effect, allowing us to use exact diagonalization to directly compute the EE. Secondly, the cylinder $S^2\times \mathbb R$ is conformally equivalent to $\mathbb R^3$ \cite{Cardy1984,Cardy1985}, such that a conformal transformation turns the EE area law in Eq.\eqref{eq:entropicF} into~\cite{Banerjee2016,Myers2010Seeing,Myers2011}
\begin{equation}\label{eq:arealaw}
S_A (\theta) = -\textrm{Tr}(\rho_A \ln \rho_A) = \frac{\alpha R}{\delta}\sin\theta - F,
\end{equation}
for an entanglement cut on the radius-$R$ sphere $S^2$ at a latitude parameterized by $\theta$, as shown in Fig.\ref{fig:illustration}. Lastly, on the fuzzy sphere, where space is continuous, we can define a cylinder renormalized entanglement entropy (cREE)~\cite{Banerjee2016} similar to the REE in Eq.~\eqref{eq:REE}:
\begin{equation}\label{eq:cREE}
\mathcal F_C (R, \theta_0) \equiv (\tan \theta \partial_{\theta} -1) S_A(\theta)|_{R,\theta_0}  .
\end{equation}
In the thermodynamic limit, where $R\sin\theta_0 \rightarrow \infty$, $\mathcal F_C(R,\theta_0)$ approaches $F$ at the IR fixed point. In this paper, leveraging the advantages of fuzzy sphere regularization, we report the first non-perturbative computation of the $F$-function for the 3D Ising CFT. Specifically, we find that $F_{\text{Ising}} \approx 0.0612(5)$, slightly smaller than the $F$-function of a real free scalar, $F_{\text{free}} = \frac{\log 2}{8} - \frac{3\zeta(3)}{16\pi^2} \approx 0.0638$ \cite{Klebanov2011}, which is consistent with the $F$-theorem. Our results are also very close to those of the $4-\epsilon$ expansion, which gives $F_{\text{Ising}} \approx 0.0610$~\cite{Giombi2015} and $F_{\text{Ising}} \approx 0.0623$ \cite{Fei2015}.

\begin{figure}[b]
    \centering
        \includegraphics[width=0.49\textwidth]{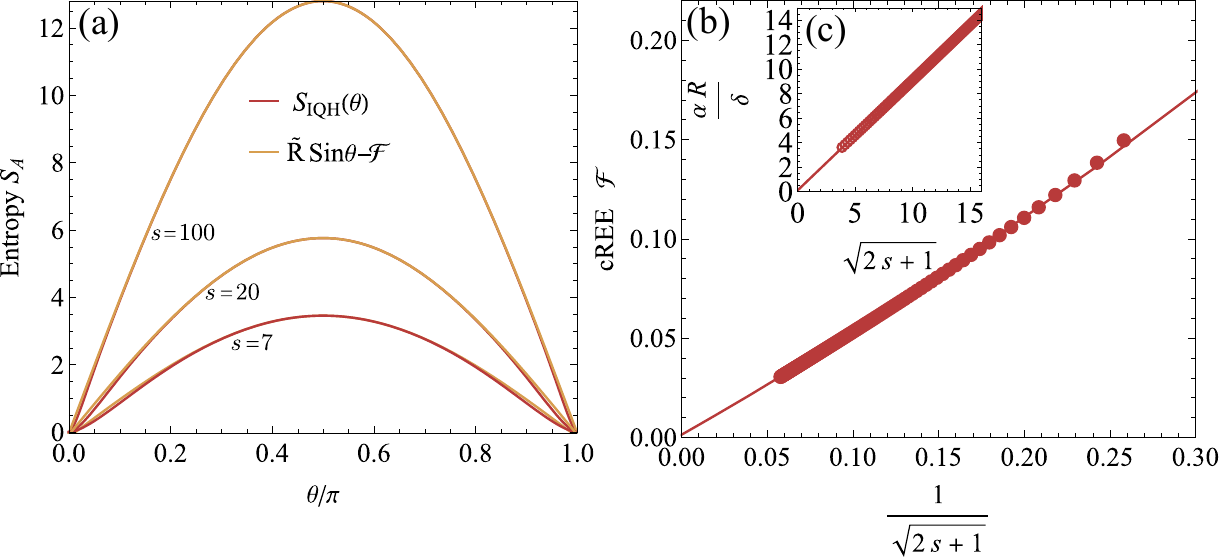}
        \caption{Results of the integer quantum Hall state. (a) EE with respect to $\theta$ for $s=100, 20, 7$. (b) The renormalized entropy versus $1/\sqrt{2s+1}$. The inset is $\alpha R/\delta$ versus $\sqrt{2s+1}$.}
        \label{fig:IQH}
\end{figure}

\emph{Fuzzy sphere and real-space EE.---}The fuzzy sphere scheme~\cite{ZHHHH2022} regularizes 3D CFTs on the continuous space-time geometry $S^2\times \mathbb R$ by investigating strongly interacting quantum mechanical models projected onto the lowest Landau level (LLL)~\cite{Ippoliti2018Half} on a sphere. For instance, to realize the 3D Ising CFT, one considers fermions moving on the sphere in the presence of a magnetic monopole with the flux $4\pi s$, and the interactions between fermions are described by: 
\begin{align} \label{eq:Hamcontinuum}
 \int d^2 \bm r_a d^2 \bm r_b \, U(\bm r_{ab})  2n^\uparrow(\bm r_a)n^\downarrow(\bm r_b) - h \int  d \bm r \, n^x(\bm r),
\end{align}
where the fermions $\bm \psi^\dag (\bm r) = (\psi^\dag_\uparrow (\bm r), \psi^\dag_\downarrow (\bm r))$ are carrying an isospin degree of freedom, and $n^{\uparrow(\downarrow)}(\bm r) = \psi^\dag_{\uparrow(\downarrow)}(\bm r) \psi_{\uparrow(\downarrow)}(\bm r)$, $n^\alpha(\bm r) = \bm{\psi}^\dag(\bm r) \sigma^a \bm{\psi}(\bm r)$ represent the spin degrees of freedom that go through the 2+1D Ising transition. Here, $U(\bm r_{ab}) = g_0 \delta(\bm r_{ab}) + g_1 \nabla^2 \delta(\bm r_{ab})$ is taken to be the same as the previous literature~\cite{ZHHHH2022}. By setting the interaction strength to be much smaller than the Landau level gap, we can project the system onto the LLL~\cite{Sphere_LL_Haldane}, $\psi_\alpha(R, \theta,\varphi)=\frac{1}{R}\sum_{m=-s}^s Y^{(s)}_{s,m}(\theta,\varphi) \hat{c}_{m,\alpha}$, where the monopole Harmonics $Y^{(s)}_{s, m}(\theta,\varphi)$~\cite{WuYangmonopole} is the single particle wavefunction of LLL. On the LLL, the sphere radius $R$ will be related to the number of LL orbitals $N=2s+1$, $R\sim \sqrt{N}$. At the half-filling of LLL, it has been shown that critical behavior of  Eq. \eqref{eq:Hamcontinuum} falls into the 3D Ising universality class \cite{ZHHHH2022}.

This paper will focus on the quantum entanglement properties of the 3D Ising criticality on the fuzzy sphere.  To incorporate the EE (Eq. \eqref{eq:arealaw}) and the cREE (Eq. \eqref{eq:cREE}), we need to calculate the continuous angular dependence of the EE on the sphere (see Fig. \ref{fig:illustration}). 
Fortunately, the transformation between the orbital-partition and real-space-partition reduced density matrix has been well-addressed before \cite{Rodriguez2012,Dubail2012,Sterdyniak2012,Zaletel2012}. 
Here we outline the general procedure.  
The real-space partition requires to account the distribution of single-particle Landau orbital relative to the entanglement cut position. 
Defining $p^{\mathcal A( \mathcal B)}$
as the weight of each Landau orbital belong to subsystem $\mathcal A(\mathcal B)$, we split the second-quantization form of electron operator as \cite{Rodriguez2012,Dubail2012,Sterdyniak2012,Zaletel2012}
\begin{equation}
    \hat c_{m,\sigma} = p^{\mathcal A}_m\hat c_{m,\sigma}(\mathcal A)+
    p^{\mathcal B}_m\hat c_{m,\sigma}(\mathcal B).
\end{equation}
In specific, by choosing subsystem $\mathcal A$ the hemisphere region $[0,\theta]$ and subsystem $\mathcal B$ the complementary region (see Fig. \ref{fig:illustration}), \(p^{\mathcal A}_m\) and $p^{\mathcal B}_m=\sqrt{1-(p^{\mathcal A}_m)^2}$ can be explicitly obtained \cite{Rodriguez2009}
\begin{equation}
    (p^{\mathcal A}_m)^2 = \int_{\mathcal A}  |Y_{s,m}^{(s)}(\theta,\varphi)|^2 
     = I_{\cos^2\left(\frac{\theta}{2}\right) }(s+m+1,s-m+1)
\end{equation}
where $I_z(a,b)$ is the regularized Beta function. Next, one can decompose the ground state wave function into $|\text{0}\rangle = \sum_{a,b} \mathcal{M}_{ab}(\theta) |a\rangle|b\rangle$
where $|a\rangle[|b\rangle]$ represent the states spanned by $c_{m,\sigma}^\dagger(\mathcal A) (c_{m,\sigma}^\dagger(\mathcal B))$. Thus, the reduced density matrix can be obtained by tracing out states in region $\mathcal B$
$$
     \rho_{\mathcal A}(\theta) = \mathrm{Tr}_{\mathcal B} |\text{0}\rangle\langle\text{0}| = \sum_{b}  \langle b|\text{0}\rangle\langle\text{0}|b\rangle,
$$
which gives the EE, $S_A(\theta)=-\textrm{Tr}( \rho_{\mathcal A}(\theta) \ln  \rho_{\mathcal A}(\theta))$.

\emph{Benchmark for the integer quantum Hall (IQH) state.--} The IQH state is a product state in the Landau orbital basis with no quantum entanglement. In the real space, the IQH state is highly entangled, whose EE is known analytically \cite{Rodriguez2009},
\begin{equation}\label{eq:IQH1}
S_{\mathrm{IQH}}(\theta) = -\sum_{n=0}^{2s} \left[\lambda_n \ln(\lambda_n) + (1-\lambda_n) \ln(1-\lambda_n) \right],    
\end{equation}
with $\lambda_n$ be a continuous function of $\theta$,
\begin{align}\label{eq:IQH2}
\lambda_n  =(2 s+1) \binom{2 s}{n} & \left(B_1(n+1,-n+2 s+1)  \right. \\ & \left. -B_{\cos ^2\left(\frac{\theta
   }{2}\right)}(n+1,-n+2 s+1)\right),     \nonumber
\end{align}
and $B_z(a,b)=\int_0^z t^{a-1} (1-t)^{b-1} dt$ is the incomplete Beta function.

Theoretically, one would expect that in the thermodynamic limit $s\rightarrow \infty$, $S_{\mathrm{IQH}}(\theta)$ follows the entanglement area law Eq.~\eqref{eq:arealaw} with $F=0$ and $R/\delta \propto \sqrt{2s+1}$. However, it is not immediately evident that the analytical results for IQH in Eq.\eqref{eq:IQH1}-\eqref{eq:IQH2} will align with this simple entanglement area law. This makes it a really interesting benchmark to look into first. As shown in Fig.\ref{fig:IQH}(a), we examine the EE of IQH for system sizes $s=7, 20, 100$. We specifically compare these results against the entanglement area law in Eq.~\eqref{eq:arealaw}, where $\tilde R(s)$ is defined as $\alpha R/\delta\equiv \frac{1}{\cos\theta} \partial_\theta S_{\mathrm{IQH}}|_{s,\theta=\pi/2}$ and $\mathcal F(s)$ as $(\tan \theta \partial\theta - 1) S_{\mathrm{IQH}} |_{s, \theta=\pi/2}$. Remarkably, the two curves—$S_{\mathrm{IQH}}(\theta)$ and $\tilde R(s) \sin\theta - \mathcal F(s)$—show a good agreement over a broad range of $\theta$ values near the equator $\theta=\pi/2$. This agreement improves progressively as system size $s$ increases.

We find that $\mathcal F(s)$ decreases with system size $s$, and $\lim_{s\rightarrow\infty} \mathcal F(s)=0$, as shown in Fig.\ref{fig:IQH}(b). We note that the cREE $\mathcal F(s)$ exhibits significant finite size effects. For instance, $\mathcal F(s=7)\approx 0.149$. This size dependence, however, should not be seen as a weakness; rather, it contains interesting physical information related to the RG flow from UV to IR, presenting intriguing avenues for further investigation. Furthermore, Fig.\ref{fig:IQH}(c) illustrates the expected behavior $\tilde R(s) \propto \sqrt{2s+1}$.

\begin{figure}[b]
    \centering
\includegraphics[width=0.23\textwidth]{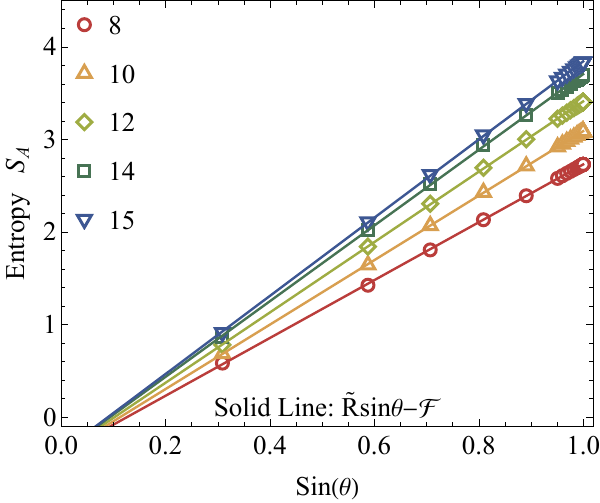}
\includegraphics[width=0.23\textwidth]{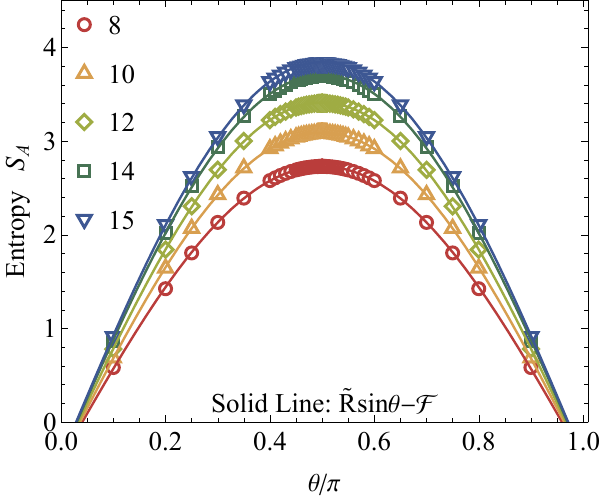}
\caption{The real-space EE of 3D Ising CFT on fuzzy sphere. The left (right) panel shows the EE versus the angle $\sin \theta$ ($\theta$) of entangling surface (see illustration in Fig. \ref{fig:illustration}), demonstrating the entanglement area law.} \label{fig:Ising}
\end{figure}

\begin{figure}[t]
    \centering
    \includegraphics[width=0.34\textwidth]{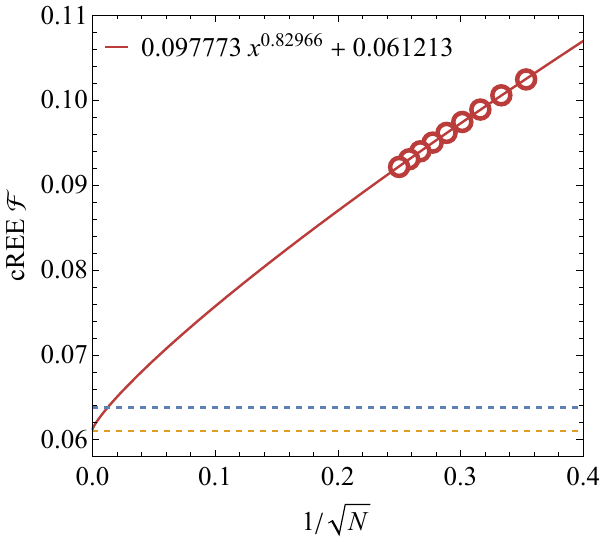}
    \caption{ Entropic $F$-function of the 3D Ising CFT on the fuzzy sphere. The finite-size scaling is based on $\mathcal F(N) = F_{\mathrm{Ising}} + \alpha (\sqrt{N})^{-0.82966}$ (see main text for details), and it gives $F_{\mathrm{Ising}} \approx 0.0612(5)$. The data points are collected for the system size $N=8,9,\cdots, 16$.
    The blue dashed line shows $F_{\mathrm{free}} \approx 0.0638 $ and orange dashed line shows 
    $F_{\mathrm{Ising}} \approx 0.0610$ from perturbation calculation \cite{Giombi2015,Fei2015}. 
}
    \label{fig:IsingF}
\end{figure}

\emph{F-function of 3D Ising CFT.--} Now we turn to the $F$-function of 3D Ising CFT. We first compute the real-space EE $S_{\mathrm{fuzzy}}$ of the fuzzy sphere Ising model, and then we subtract the IQH contribution $S_{\mathrm{IQH}}$, $S_{\mathrm{Ising}}(\theta) =S_{\mathrm{fuzzy}}(\theta)-S_{\mathrm{IQH}}(\theta) $. Physically, this subtraction can be viewed as a subtraction of the contribution from the UV regularator--IQH, the state formed by the charge degree of freedom in the entire phase diagram. Different from the IQH where the analytical expression of EE is known, the EE of fuzzy sphere Ising model can only be computed numerically. Therefore, we have to numerically compute EE at specific angles $\theta$, although any value of $\theta$ is accessible in our computation.    Fig. \ref{fig:Ising} shows $S_{\mathrm{Ising}}(\theta)$ at representative angles of different system sizes, and we find they follow the entanglement area law nicely. This is sharply different from the non-commutative field theory on the fuzzy sphere, where the entanglement area law is violated~\cite{Karczmarek2014Entanglement,Sabella-Garnier2015Mutual}.

Similar to the case of IQH, we can compute the cREE by taking derivatives of $S_{\mathrm{Ising}}(\theta)$ at the equator $\theta=\pi/2$, with the difference that we need to compute the derivatives numerically. In practice, we choose two different angles, $\theta=0.5\pi$ and $\theta=0.499\pi$ (See Supple. Mat. Sec. B \cite{sm}), and fit $S_{\mathrm{Ising}}$ with the entanglement area law $\tilde R(N) \sin(\theta) -\mathcal F(N)$. Fig.~\ref{fig:IsingF} shows the cREE $\mathcal F(N)$ for different system sizes $N=8,9,\cdots, 16$. Similar to the IQH, cREE $\mathcal F(N)$ shows considerable size dependence, and it monotonically decreases with the system sizes. This size dependence is again a manifestation of RG flow. There are two sources for the size dependence, i.e., the perturbation from i) RG irrelevant operators and ii) curvature effect. The contribution from an irrelevant operator $O$ will be $ R^{\Delta_O-3} \sim N^{(\Delta_O-3)/2} $, while the curvature perturbation will lead to corrections like $R^{-1}$, $R^{-3}$, etc. For the Ising CFT, the leading irrelevant operator $\epsilon'$ has $\Delta_{\epsilon'}\approx 3.82966$, so it will contribute a finite size correction $R^{-0.82966}$. We fit the finite size correction according to $\mathcal F(N) = F_{\mathrm{Ising}} + \alpha (\sqrt{N})^{-0.82966}$, and we obtain $F_{\mathrm{Ising}} \approx 0.0612(5)$. The error is estimated using the standard deviation of results obtained from different fitting ranges, where the fitting range is defined as $[N_{\text{min}}, 16]$, with $N_{\text{min}}$ varying from 8 to 14. Our estimate $F_{\mathrm{Ising}}$ is slightly smaller than the $F$-function of a real free scalar, $F_{\mathrm{free}} = \frac{\log 2}{8} - \frac{3\zeta(3)} {16\pi^2}  \approx 0.0638 $ \cite{Klebanov2011}, which is consistent with the $F$-theorem, $F_{UV}>F_{IR}$. We also note that $4-\epsilon$-expansion gives $F_{\mathrm{Ising}} \approx 0.0610$ and $F_{\mathrm{Ising}} \approx 0.0622$ \cite{Giombi2015,Fei2015}, which are very close to our estimates.

We remark that, in principle, we can include higher-order corrections to fit the $F$-function. The next order will be $(\sqrt{N})^{-1}$, which is close to the leading order $(\sqrt{N})^{-0.82966}$. Practically, we find that including the subleading term does not significantly change the results. However, it can easily lead to overfitting, so we just include the leading order for the finite size analysis.

\emph{Orbital space entanglement.---}The LLL orbital wavefunctions are Gaussian-localized at different latitudes of the sphere. Intuitively, one might expect that the entanglement entropy in orbital space will be qualitatively similar to that in real space. Therefore, it is natural to inquire whether one can also extract the $F$-function from the orbital space entanglement, a question on which we will briefly comment. In the orbital partition, the subsystem size can only be tuned discretely, in contrast to the real-space partition. We investigated the EE between the $m>0$ and $m<0$ LL orbitals, finding that they indeed follow the entanglement area law $S\propto \sqrt{N}$. However, the subleading term is 0.159(1), which is much larger than the $F$-function of the free scalar. Additionally, we performed an analysis following Ref.\cite{PhysRevLett.98.060401} (see Sec. C of Supplementary Material \cite{sm}), but it did not yield an accurate $F$-function either.

\emph{Discussion.--}We have conducted a comprehensive investigation into the entanglement entropy (EE) of 3D Ising CFT on the fuzzy sphere. We find that EE follows the entanglement area law with a subleading term corresponding to the $F$-function, an RG monotonic quantity that measures the number of degrees of freedom of the QFT. Our non-perturbative calculation yields $F_{\text{Ising}} \approx 0.0612(5)$, which is slightly smaller than the real free scalar $F_{\text{free}} \approx 0.0638$ \cite{Klebanov2011}, consistent with the $F$-theorem. Our work represents the first non-perturbative computation for the $F$-function of the paradigmatic 3D Ising CFT. The same scheme can be applied to other universality classes, such as Wilson-Fisher $O(N)$ criticalities and critical gauge theories, and will provide many new insights into understanding the landscape of 3D CFTs.  

Besides the subleading entropic $F$-function, it is also worth highlighting the entanglement area law in our fuzzy sphere Ising model. It has been found that in a non-commutative field theory defined on the fuzzy sphere \cite{Karczmarek2014Entanglement,Sabella-Garnier2015Mutual}, the entanglement area law is violated, which was interpreted as a consequence of UV-IR mixing. Our observation of the entanglement area law can serve as further evidence that a fully local QFT is realized in our fuzzy sphere model. Delving into this may provide new perspectives on a century-old dream, originating from Heisenberg, namely, regularizing QFTs with the non-commutative geometry.

In the current work, we primarily focus on the EE of the ground state. An interesting future direction would be to explore the EE of excited states, as well as other entanglement measures such as entanglement spectra. In the Supplementary Material, Section 
 A.2 \cite{sm}, we investigate the EEs of excited primary states $|\sigma\rangle$ and $|\epsilon\rangle$, finding that they also adhere to the entanglement area law. It would be intriguing to examine whether higher excited states of CFT also follow the entanglement area law and whether other universal quantities are encoded within it.

\begin{acknowledgments}
\textit{Acknowledgments.---} We thank Max Metlitski and Rob Myers for stimulating discussions. 
LDH and WZ were supported by National Natural Science Foundation of China (No.~92165102) and National key R$\&$D program (No. 2022YFA1402204).  
Research at Perimeter Institute is supported in part by the Government of Canada through the Department of Innovation, Science and Industry Canada and by the Province of Ontario through the Ministry of Colleges and Universities.
\end{acknowledgments}

\bibliography{ref}

\clearpage
\appendix
\begin{widetext}

\begin{center}
\textbf{Supplementary materials for `` Entropic $F$-function of 3D Ising conformal field theory via the fuzzy sphere regularization ''}    
\end{center}
\setcounter{subsection}{0}
\setcounter{equation}{0}
\setcounter{figure}{0}
\renewcommand{\theequation}{S\arabic{equation}}
\renewcommand{\thefigure}{S\arabic{figure}}
\renewcommand{\thetable}{S\arabic{table}}
\setcounter{table}{0}
\appendix
In this supplementary material, we will demonstrate more details to support the discussion in the main text. In Sec. A, we present the entanglement entropy for the ferromagnetic and paramagnetic phase, and the entanglement entropy for the low-energy excited states at the phase transition point.
In Sec. B, we show how to deal with discrete derivative. 
In Sec. C, we show the orbital entanglement entropy.

\section{A. More detailed results}
\subsection{1. $h$-dependence of $F$ term}
In the main text, we focus on the result of $F$ function in the critical point $h_c=3.16$. Here we explore the  cylinder
 renormalized entanglement entropy (cREE) in ferromagnetic or paramagnetic phases. The ferromagnetic phase has a non-trivial $F=-\ln 2$, which is a consequence of two-fold degeneracy from the spontaneous $\mathbb Z_2$ symmetry breaking. The paramagnetic phase, on the other hand has $F=0$ because the ground state is a trivially gapped with no degeneracy.  In Fig. \ref{smfig:IsingF}, we have shown the cREE in different $h$ from the ferromagnetic phase to the paramagnetic phase. We can find that the cREE  converges to $-\ln 2$ and $0$ in the ferromagnetic phase to the paramagnetic phase respectively. In the intermediate region, the cREE develops a peak  near the critical point, which indicates the presence of a critical point. 
\begin{figure}[b]
    \centering
    \includegraphics[width=0.36\textwidth]{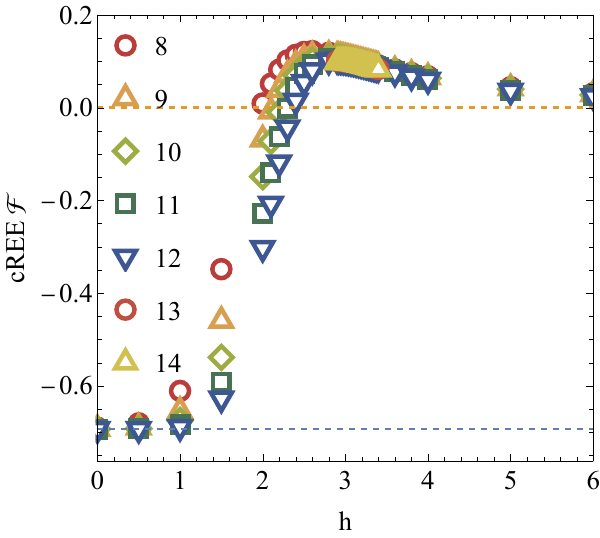}
    \caption{ cREE of the 3D Ising CFT in different $h$.  In the deep ferromagnetic phase($h\sim 0$), the cREE collapses to the expected value $-\ln2$(blue dashed line). In the deep paramagnetic phase ($h\gg h_c$), the cREE tends to zero. 
    }
    \label{smfig:IsingF}
\end{figure}

\subsection{2. Entanglement Entropy of $|\hat \sigma\rangle$ and $|\hat \epsilon\rangle$ }
\begin{figure}
    \centering
    \includegraphics[width=0.24\textwidth]{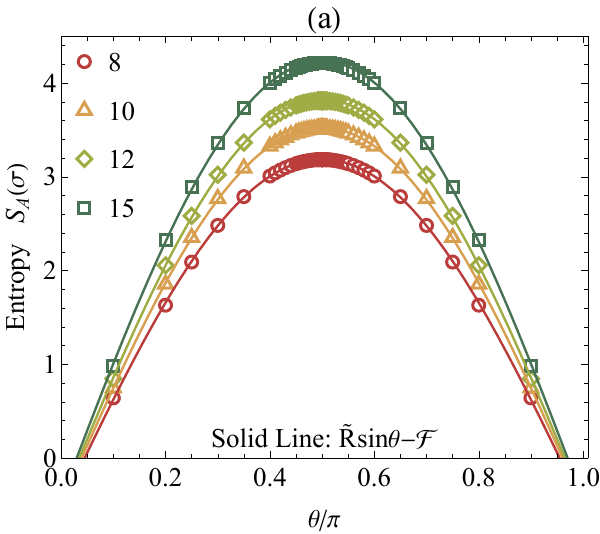}
    \includegraphics[width=0.24\textwidth]{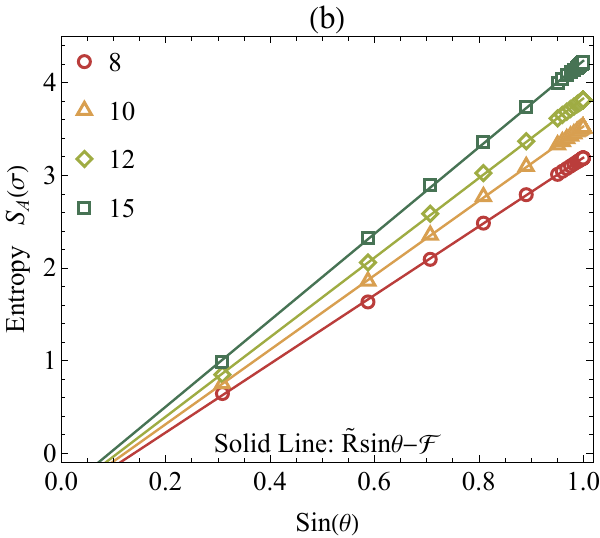}
    \includegraphics[width=0.24\textwidth]{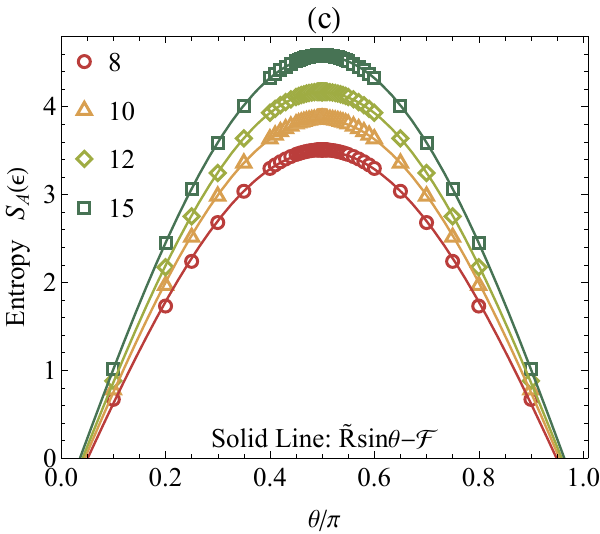}
    \includegraphics[width=0.24\textwidth]{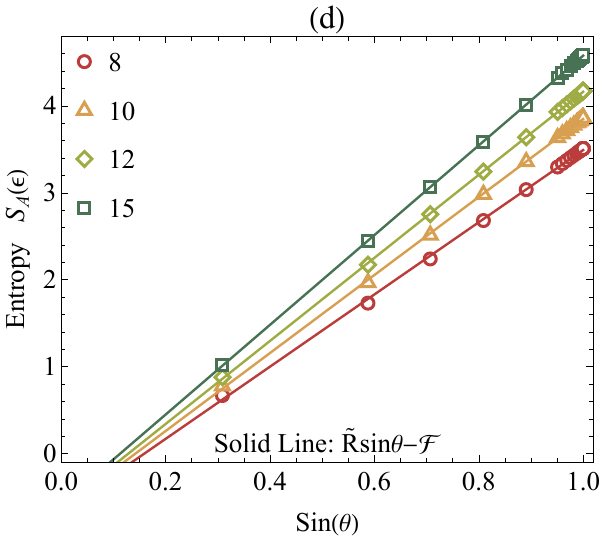}
    \caption{ Entanglement entropy's area law for primary fields. (a-b) Entanglement entropy for $\sigma$, where open markers represent numerical results, and solid lines depict fitting results near the equator using the area law formula $S=\tilde{R}\sin\theta-\mathcal{F}$; (c-d) Similar results for $\epsilon$.}
    \label{smfig:sigmaepsilon}
\end{figure}
Having concluded our discussion on the ground state entanglement, we now shift our focus to the entanglement properties of primary fields at the critical point. In this section, we take the example of the lowest excitations in two distinct $\mathbb Z_2$ sectors, namely $\sigma$ and $\epsilon$. To assess whether they satisfy an area law, we employ the function $S = \tilde{R} \sin\theta - \mathcal{F}$ for fitting numerical results near the equator. Subsequently, we compare the fitted results with the numerical results to verify their adherence to the area law. As depicted in Fig.\ref{smfig:sigmaepsilon}, both $\sigma$ and $\epsilon$ exhibit excellent compliance with the area law for their entanglement entropy. This indicates that at the critical point, low-energy excitations of primary fields also adhere to the area law. However, it is unclear about the physical meaning of the subleading term.

\section{B. Discrete derivative and convergence.}
In the main text, extracting $F$ requires the entanglement entropy's derivative concerning the angle near the equator, 
\begin{equation}
    \mathcal F_C (R, \theta_0) \equiv (\tan \theta \partial_{\theta} -1) S_A(\theta)|_{R,\theta_0}  .
\end{equation}
In practice, we discretize the derivative into differences
\begin{equation}
\mathcal F_C (R, \theta) \approx \left. \frac{S(\frac{\pi}{2})-S(\theta)}{1-\sin\theta}-S\left(\frac{\pi}{2}\right)\right|_{\theta=\frac{\pi}{2}-\alpha}  .
\end{equation}
Here, $\alpha$ represents the angular deviation from the equator. In this section, we vary the angle $\alpha$ within a certain range to observe the convergence of the discrete derivative. As illustrated in Fig. \ref{smfig:ddconvergence}, we consider angle changes in the range $\alpha \in [0.001\pi, 0.05\pi]$ and compare the obtained results with $\mathcal F(0.4999\pi)$. From the results in the figure, it is evident that for $\alpha \le 0.01\pi$, the error in the derivative is only on the order of $10^{-5}$, making it entirely reliable.
\begin{figure}
    \centering
    \includegraphics[width=0.64\textwidth]{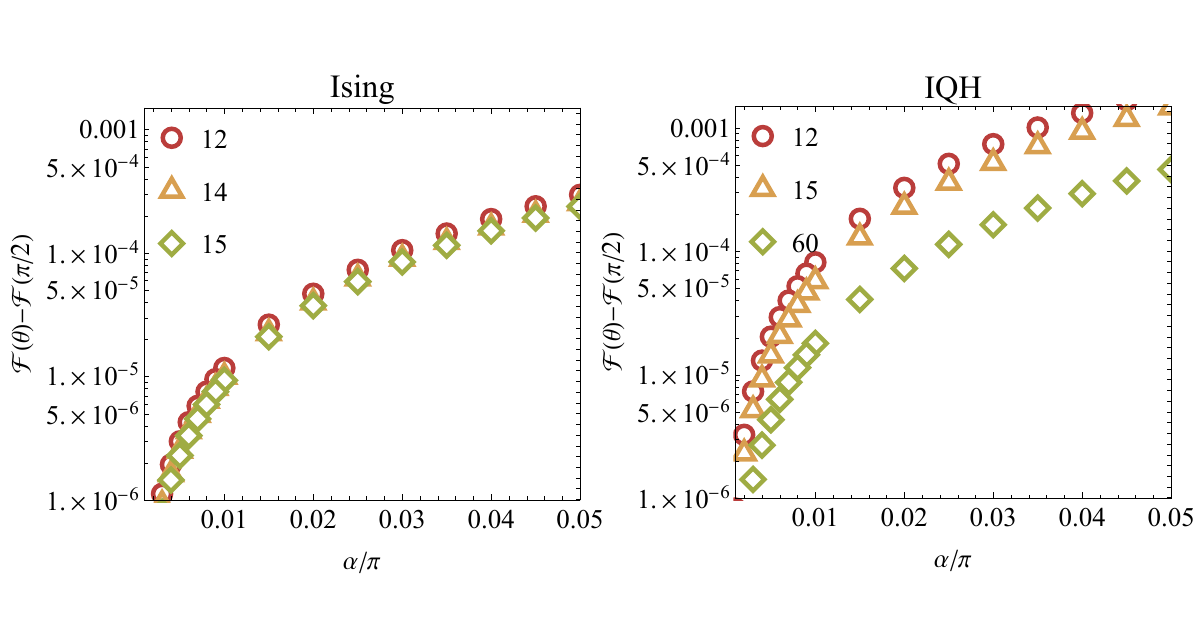}
    \caption{Convergence of the discrete derivative near the equator: (Left) Results for the Ising critical point; (Right) Results for the Integer Quantum Hall (IQH) system. Different points in the figure represent various system sizes, denoted by $N=12,15,60$.}
    \label{smfig:ddconvergence}
\end{figure}

\section{C. Orbital Bipartition}
In this section, we will discuss the properties of entanglement entropy in the context of orbital space bipartition. Orbital bipartition is much simpler compared to real-space bipartition, allowing for the inclusion of larger system sizes. To calculate the entanglement entropy of orbital space bipartition, we first need to divide the orbital space into two parts, denoted as $A$ and $B$, defined as $A=\{m=-s,-s+1,\cdots,-s+l_A-1\}$ and $B=\{m=-s+l_A,-s+l_A+1,\cdots,s\}$, where $l_A$ is the size of subsystem A. The first property we need to check is the area law. In Fig. \ref{smfig:OEE}(a), we have plotted the entanglement entropy of different subsystem sizes $l_A$. It is not easy to discern the area law from the graph because in orbital space, we cannot define the ``boundary'' of subsystem A. However, we can use $\sqrt{l_A}$ to characterize the ``boundary'' area of subsystem A. Therefore, we plot the entanglement entropy of the subsystem against $\sqrt{l_A}$ in Fig. \ref{smfig:OEE}(b). It can be observed that, away from the center (i.e., for symmetric bipartition), the entanglement entropy approximately follows a linear relation with $\sqrt{l_A}$. However, this behavior is severely disrupted at the central position due to finite-size effects. 
We can also directly confirm the area law by investigating the entanglement entropy at different sizes under symmetric bipartition, denoted as $S_{l_A=N/2}$. In this scenario, $A=\{-s,-s+1,\cdots,-1/2\}$ and $B=\{1/2,3/2,\cdots, s\}$. As shown Fig. \ref{smfig:OEE}(c), the results satisfactorily adhere to the area law. Through linear fitting, we extract a subleading term 0.159(1), significantly exceeding the theoretical value for a free scalar. Therefore, a direct extrapolation of orbital entanglement entropy does not give the correct $F$-function.

We can also perform the scheme proposed in Ref. \cite{PhysRevLett.98.060401}, which was originally used to extract topological entanglement entropy of fractional quantum Hall state. For each orbital $l_A$, we first extrapolate its thermodynamic limit EE $S_{l_A}(\infty)$, $S_{l_A}(N)\approx S_{l_A}(\infty)+c_1/N+c_2/N^2$, and then examine how $S_{l_A}(\infty)$ scales with $l_A$. 
In Fig. \ref{smfig:TOEE}, the left part shows the results of the extrapolation fitting, while the right part displays the entanglement entropy in the thermodynamic limit plotted against $\sqrt{l_A}$. It is evident that $S_{l_A}(N\rightarrow\infty)$ satisfies the area law. Through linear fitting, we obtain the a subleading term  $0.087(2)$, larger than the $F$-function of free scalar. So this scheme will not give correct $F$ either.
\begin{figure}
    \centering
    \includegraphics[width=0.9\textwidth]{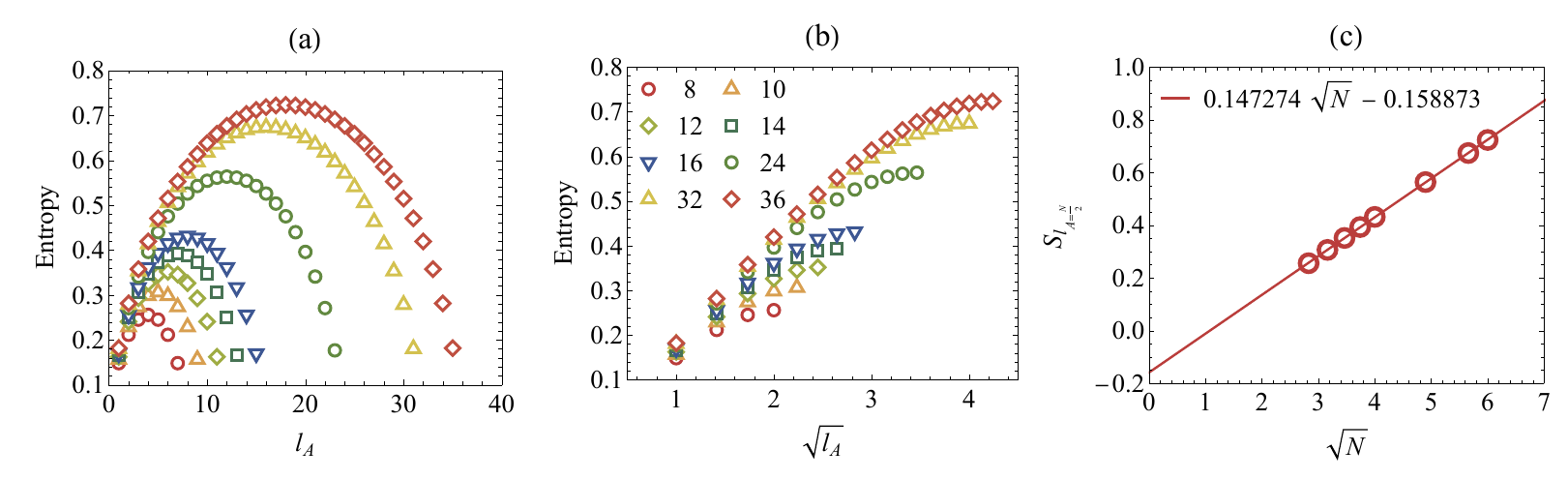}
    \caption{ Entanglement entropy behavior in orbital space bipartition. (a) The entanglement entropy for different subsystem sizes $l_A$ is plotted.(b) The entanglement entropy plotted against $\sqrt{l_A}$, emphasizing a linear relationship away from the center but disruption at the central position due to finite-size effects. (c) The entanglement entropy for symmetric bipartition(i.e. equator) at different sizes satisfies the area law, and the subleading term extracted through linear fitting is 0.159(1).  The system size from $N=8$ to $N=36$. }
    \label{smfig:OEE}
\end{figure}

\begin{figure}
    \centering
\includegraphics[width=0.7\textwidth]{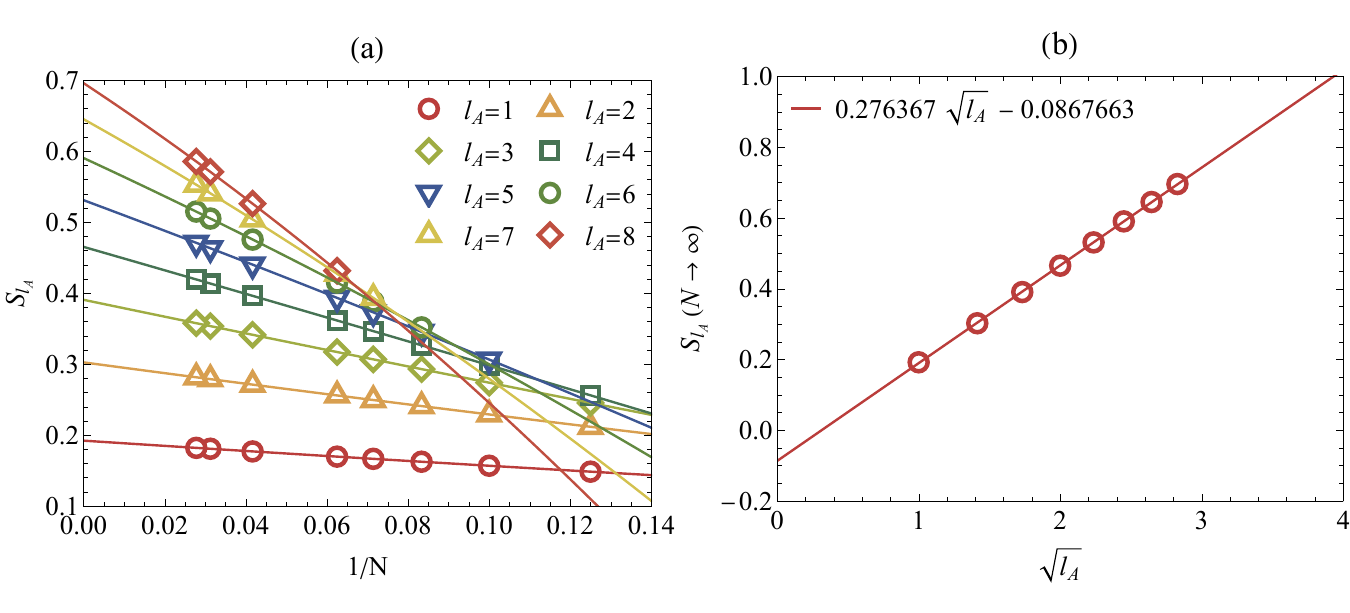}
    \caption{ (q) The extrapolation results for different subsystem sizes $l_A$. (b) Through linear fitting, the corresponding cREE for orbital space bipartition is determined as $\mathcal{F}\approx 0.087\pm0.002$. }
    \label{smfig:TOEE}
\end{figure}

\end{widetext}
\end{document}